\documentclass[a4paper,11pt]{article}

\usepackage{jheppub}
\usepackage{multicol,enumerate}
\usepackage{multirow,amsmath,amssymb}
\usepackage{subfigure}
\usepackage{slashed} 
\usepackage{color} 
\newcommand{\met}{\ensuremath{\slashed{E}_T}}

\newcommand{\fbinv} {\mbox{\ensuremath{~\text{fb}^\text{$-$1}}}}

\newcommand{\pythia}{{\sc Pythia}}
\newcommand{\delphes}{{\sc Delphes}}

\newcommand{\mgme}{{\sc MadGraph/MadEvent}}

\newcommand{\B}{\cal{B}}

\title{VBS ${\rm W}^\pm {\rm W}^\pm {\rm H}$ production at the HL-LHC and a 100 TeV pp-collider}

\author{Christoph Englert$^a$,}
\author{Qiang Li$^{b,c}$,}
\author{Michael Spannowsky$^d$,} 
\author{Mengmeng Wang$^b$,}
\author{Lei Wang$^b$}

\affiliation{$^a$SUPA, School of Physics and Astronomy, University of Glasgow, Glasgow, G12 8QQ, UK}
\affiliation{$^b$Department of Physics and State Key Laboratory of Nuclear Physics and Technology,\\
Peking University, Beijing, 100871, China}
\affiliation{$^c$CAS Center for Excellence in Particle Physics, Beijing 100049, China}
\affiliation{$^d$Institute for Particle Physics Phenomenlogy, Department of Physics,\\Durham University, DH1 3LE, UK}

\emailAdd{christoph.englert@glasgow.ac.uk}
\emailAdd{mengmeng.wang@cern.ch}
\emailAdd{qliphy0@pku.edu.cn}
\emailAdd{michael.spannowsky@durham.ac.uk}
\emailAdd{melodyphysics@gmail.com}

\abstract{
${\rm W}^\pm {\rm W}^\pm {\rm H}$ production at hadron colliders through vector boson scattering is a so far unconsidered process, which leads to a clean signature of two same-sign charged leptons and two widely separated jets. This process is sensitive to the ${\rm HHH}$ and ${\rm WWHH}$ couplings and any deviation of these couplings from their SM predictions serves as direct evidence of new physics beyond the SM. In this paper we perform a Monte Carlo study of this process for the $\sqrt{s}=14$~TeV LHC and a $100$~TeV pp-collider, and provide projections of the constraints on the triple-Higgs and ${\rm WWHH}$ quartic couplings for these environments. In particular, we consider the impact of pileup on the expected sensitivity in this channel. Our analysis demonstrates that although the sensitivity to the ${\rm HHH}$ coupling is rather low, the ${\rm WWHH}$ coupling can be constrained in this channel within $\sim 100\%$ and $\sim 20\%$ at 95\% confidence level around the SM prediction at the HL-LHC and a 100 TeV pp-collider, respectively.}

\date{\Date}


\begin{document}
\maketitle
\flushbottom

\section{Introduction}
\label{intr}

\qquad After the discovery of the 125 GeV Higgs-like boson~\cite{FGianotti,JIncandela,plb:2012gu,plb:2012gk}, one of the primary goals of present collider phenomenology is to formulate ways to pave the way to a better understanding of the mechanism of electroweak symmetry breaking (EWSB). In particular, the trilinear Higgs ${\rm HHH}$ and the quartic ${\rm VVHH}$ vertices (with ${\rm V}$ representing the ${\rm W}$ and ${\rm Z}$ vector bosons) are key parameters, which are also directly linked to radiative instability of the TeV scale~\cite{Veltman:1980mj}, as well as to a potential radiative nature of EWSB~\cite{Coleman:1973jx,Agashe:2004rs,Contino:2003ve}.

In the Standard Model (SM), the ${\rm WWHH}$ coupling is determined by electroweak gauge invariance, which enforces $g_{{\rm WWHH}}=e^2/(2s_w^2)$, with $s_w$ denoting the sine of the Weinberg angle and $e$ the electric charge, respectively. Any deviation from this value indicates the necessary existence of new physics beyond the SM, as a departure from the gauge-relations directly induces (perturbative) unitarity violation \cite{Lee:1977eg}, unless new resonant states mend the dangerous growth of the $\rm{WW\to HH}$ amplitude. Models with extra dimensions and their holographic interpretation in terms of composite theories are well-known examples of how such coupling modifications can appear in the low energy formulation of strongly interacting scenarios (for a recent review see \cite{Csaki:2015hcd}). Modifications of unitarity sum rules can be used to predict some properties of new composite states~\cite{Birkedal:2004au,Englert:2015oga,Thamm:2015zwa}. In such scenarios, only measuring the trilinear gauge couplings is not necessarily indicative of the quartic gauge couplings in the low energy effective field theory (EFT), as new states are crucial to enforce $d> 4$ gauge invariance in the dual holographic picture. Bearing scenarios like this in mind, there is motivation to isolate the sensitivity to the quartic couplings in collider processes.\footnote{Integrating out extra states leads to a plethora of modified couplings that can be investigated in a global non-linear SM EFT fits~\cite{Ellis:2014jta,Falkowski:2015jaa,Englert:2015hrx,Falkowski:2016cxu,Butter:2016cvz}; the focus of our work is to discuss the sensitivity of a particular process that could be exploited in this direction as well.} 

Aiming to probe the ${\rm VVHH}$ couplings at hadron colliders, one usually thinks of exploiting processes with two final state Higgs bosons. This final state has been investigated in~Refs.~\cite{Dolan:2013rja,Dolan:2015zja} (see also \cite{Bishara:2016kjn}), which have shown that focussing on the vector boson scattering (VBS) component of ${\rm{HH}}$+2~jets production can in principle constrain the quartic gauge-Higgs coupling within $\sim50\%$ around the SM prediction. One of the shortcomings of such an analysis is that {\emph{all}} ${\rm{VVHH}}$ couplings contribute coherently. Systematically distinguishing between the contributing couplings as would be required to phenomenologically reverse-engineer, e.g., the Veltman condition is not possible, in particular given the low statistical yield.

In this paper, we focus on a so far unconsidered process, $pp\to {\rm W}^\pm {\rm W}^\pm {\rm H}$+2 jets, with $W^\pm$ decaying to leptons, which is predominantly sensitive to the ${\rm WWHH}$ coupling {\it exclusively} as it does not involve the ${\rm ZZHH}$ coupling at leading order. This way, a successful analysis of this final state at present or future hadron colliders will not only provide additional information to a $\kappa$-framework analysis \cite{Heinemeyer:2013tqa} (which we will limit ourselves to in this first study), but is also likely to provide complementary information for a more comprehensive SM-EFT analysis (in particular by accessing different kinematical regimes than final states with on-shell Higgs bosons \cite{Dolan:2013rja,Dolan:2015zja}). Furthermore, it provides a relatively clean signal of two same-sign leptons and two VBS jets, analogous to the standard VBS paradigm \cite{Rainwater:1999sd}. However, due to the small production rates of this process at the current energy frontier of the LHC, as well as relatively large expected backgrounds, one must go beyond the current LHC scope to Higher Luminosity (HL) and increased collision energy. 

The so called HL-LHC is designed to reach the LHC design energy of $14$~TeV and will include upgrades to the LHC accelerator and detector environments, allowing the machine to eventually take around $3000\fbinv$ of data~\cite{HLLHC}. Another option, which has received considerable interest, is a 100 TeV pp-collider~\cite{Benedikt, CEPC, Mangano:2016}. The large  statistics that both options can accumulate will allow us to also access rare processes (including the one we are interested in) and set constraints on their potential deviations from the SM. 

In this work we provide a first detailed MC feasibility study of measuring VBS ${\rm W}^\pm {\rm W}^\pm {\rm H}$ production, and probing the quartic coupling of ${\rm WWHH}$, at the HL-LHC and 100 TeV pp-collider. Our work takes into account the effects from parton showering, detector simulation, as well as pileup; we also comment on the sensitivity of this process to the trilinear Higgs coupling. The work is organised as follows: We describe the framework of our simulation studies in Secs.~\ref{Sim} and~\ref{Sim100}, and present the numerical results in Secs.~\ref{ana} and~\ref{ana100}, for the 14 TeV HL-LHC and 100 TeV pp-collider, respectively. We present our conclusions in Sec.~\ref{talk}.
\begin{figure}[!t]
  \centering
  \includegraphics[width=0.78\textwidth]{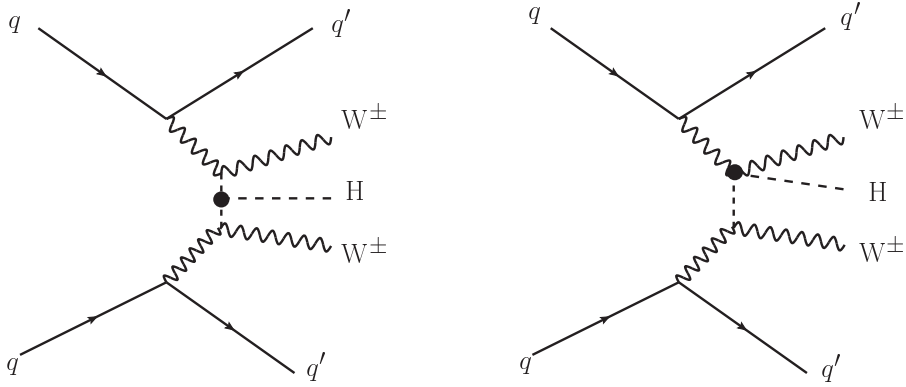}
  \caption{\label{fig:fmd} Representative Feynman diagrams for VBS same-sign
${\rm W}^\pm {\rm W}^\pm {\rm H}$ productions at the LHC, which involve the ${\rm HHH}$ and
${\rm WWHH}$ vertices. }
\end{figure}
   
\section{VBS ${\rm W}^\pm {\rm W}^\pm {\rm H}$ production at the 14 TeV LHC}
\subsection{Event Simulation and Selection}
\label{Sim}

\begin{figure}[!b]
  \includegraphics[width=0.5\textwidth]{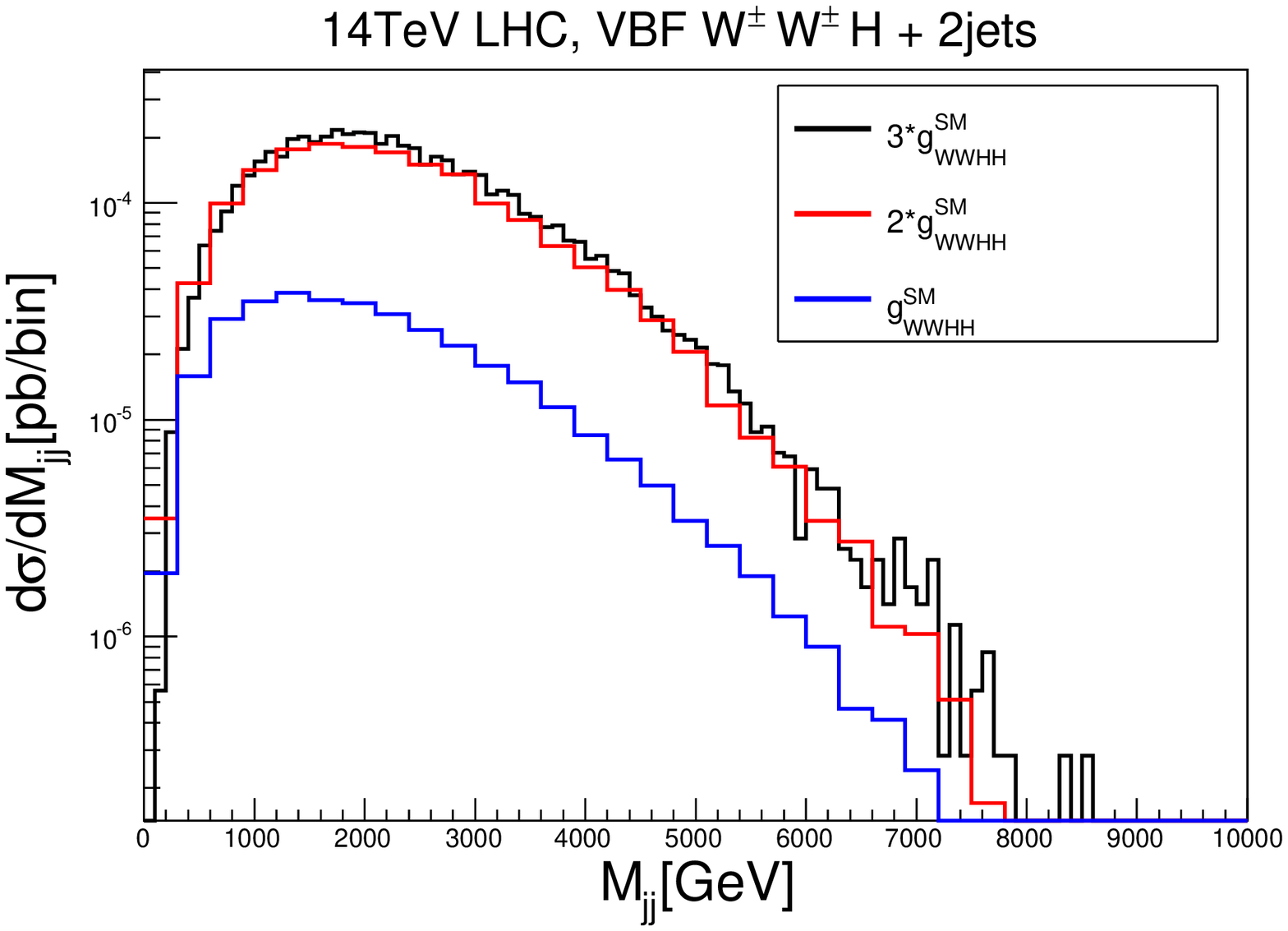}
  \hfill
  \includegraphics[width=0.5\textwidth]{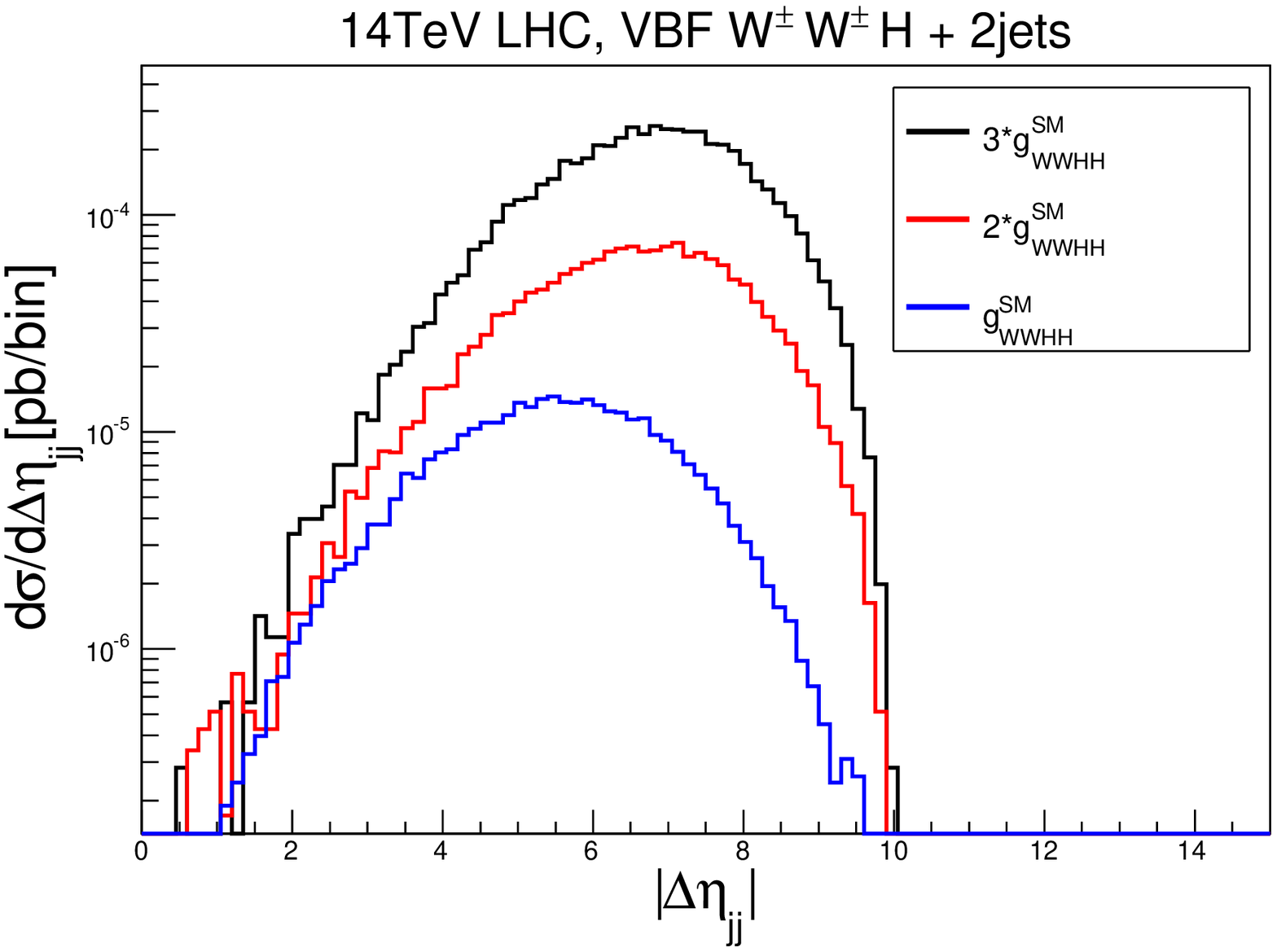}
  \caption{\label{fig:parton} ${\rm M}_{jj}$ and $|\Delta\eta_{jj}|$ distributions for ${\rm W}^\pm {\rm W}^\pm {\rm H}$ productions at the 14~TeV LHC, in the SM or varied $g_{\rm WWHH}$ cases, at parton level, with default parton level setting. }
\end{figure}

\qquad The characteristic signal that we are interested in contains two well-identified leptons (electrons $e$, or muons $\mu$) with same charge, in association with 2 VBS jets and 2 b-tagged jets. In Fig.~\ref{fig:fmd}, we show representative Feynman diagrams contributing to the VBS ${\rm W}^\pm {\rm W}^\pm {\rm H}+jj$ production at the LHC. We plot the VBS jets' invariant mass ${\rm M}_{jj}$ and pseudo-rapidity separation $|\Delta\eta_{jj}|$ at parton level, in the SM and also the cases of varied $g_{\rm WWHH}$ in Fig.~\ref{fig:parton}. As expected~\cite{Rainwater:1999sd}, the VBS-type topology leads to a sizable rapidity gap between the forward tagging jets with all weak boson-associated decay products focussed in the central region of the detector. This can be used to suppress the expected backgrounds. As can be seen from Fig.~\ref{fig:parton}, not only the total normalization of signal depends on the value of the quartic coupling, but also VBS ${\rm W}^\pm {\rm W}^\pm {\rm H}$ production tends to have harder $\rm{M}_{jj}$ and, consequently, more separated $|\Delta\eta_{jj}|$ distributions.

We follow the Snowmass Energy Frontier studies~\cite{Anderson:2013kxz,Avetisyan:2013onh,snowmassbkg} for our signal and background simulations. We take existing samples directly from Snowmass~\cite{Anderson:2013kxz,Avetisyan:2013onh, snowmassbkg}, including $t\bar{t}$, $t\bar{t}+\B$ ($\B=\gamma, {\rm W}, {\rm Z}$ or ${\rm H}$), $\B$+jets and single top. These MC samples are generated with \mgme~\cite{MadGraph:2012}, interfaced with \pythia~6~\cite{Sjostrand:2003wg} for parton showering and hadronization,  and \delphes~version 3~\cite{deFavereau:2013fsa} for detector simulation with the so-called `Combined Snowmass Detector' configuration~\cite{Anderson:2013kxz}. In Delphes, we consider no pileup (No-PU) and mean 50 pileup (PU50) scenarios at the 14~TeV LHC, and no pileup (No-PU) and 140 pileup (PU140) for the future 100~TeV pp-collider option, owing to the larger expected pileup contribution when moving from 14 to 100 TeV collisions.

It is worth mentioning that the b-tagging efficiency is rather low in the Snowmass configuration, around $20\% - 30\%$, when the b jets' transverse momentum is around $30$~GeV~\cite{Anderson:2013kxz}. However, the b-tagging efficiency could reach 70\% when the b jets' $P_T$ extends to 100~GeV.  Consequently, to enlarge the signal selection efficiency, both 2 b-tagged and 1 b-tagged jet categories should be considered, see below.

For the samples not included in the Snowmass studies, i.e. our signal VBS ${\rm W}^\pm {\rm W}^\pm {\rm H}$+2 jets and background ${\rm W}^\pm {\rm W}^\pm$ + QCD jets and VBS ${\rm W}^\pm {\rm W}^\pm$ (or ${\rm WZ}$) + jets, we produce them exactly following the description above. Finally, the analysis is based on the ExRootAnalysis~\cite{ExRootAnalysis} and ROOT~\cite{root}  packages. 

The Snowmass samples have associated NLO QCD weight factors at event level and  
thus are normalized beyond LO~\cite{Anderson:2013kxz}.  For $t\bar{t}$ process, we further apply a reweighting factor related to the most accurate prediction of Ref.~\cite{TTNNLO}.  The theoretical uncertainties at the 14~TeV LHC, are at around $5\%$, $15\%$ and $5\%$ level,  for $t\bar{t}$~\cite{TTNNLO}, $t\bar{t}H$~\cite{Heinemeyer:2013tqa} and single top processes~\cite{Kidonakis:2016smr}, respectively. We therefore assume a $20\%$ overall uncertainty on the background yields to compare with the nominal results without such systematic included, as will be shown below.
  
In our selection we require exactly 2 isolated leptons with identical charge, in addition to 2 VBS jets as well as 2 jets with a ``b-tag'' as defined below. We apply the following cuts:
\begin{enumerate}[1.)]
\item require exactly 2 leptons with same-sign charge with $P_{T\,l} \geq 20$~GeV, $|\eta_l|<2.5$ and $R_{ll} = \sqrt{\Delta\eta^2_{ll}+\Delta\phi^2_{ll}}>0.4$,  
\item require at least 4 jets. Among those we require that there is at least 1 b-tagged jet with $P_{T\,b} \geq 25$~GeV, $|\eta_b|<2.5$, and at least 2 non b-tagged jets, with $P_{T\,j} \geq 25$~GeV, $|\eta_j|<4.7$.
\begin{enumerate}[a.)]
\item If there are 2 b-tagged jets, we choose 2 VBS jets as the leading 2 non-b jets,
\item if there is only 1 b-tagged jet, we loop over the leading 3 non b-tagged jets, select the 2 VBS jets on the basis of the largest invariant mass ${\rm M}_{jj}$, and then choose the remaining jet (with additional selection $|\eta|<2.5$) to be combined with the b-tagged jet to reconstruct Higgs (we will label this with ``b'' in the following although there might not be a positive tag).
\end{enumerate} 
\item We furthermore impose $R_{bb,bj,bl}>0.4$, and $R_{jj,jl}>0.4$,  
\item and require a significant amount of missing energy \met $>30$~GeV,
\item require $|M_{ll}-M_{\rm Z}|>15$~GeV for same flavor lepton category, to suppress Drell-Yan backgrounds,
\item require $M_{ll}>50$~GeV to suppress soft lepton contributions from heavy flavor decays in W+jets and top-quark backgrounds,
\item and impose compatibility with the Higgs mass $|{\rm M}_{bb}-{\rm M}_{\rm H}|<20$~GeV,
\item We then focus on two different final cut scenarios for comparison, to further enhance the VBS signal:
\begin{enumerate}[(A)]
\item $|\Delta\eta_{jj}|>5$ and ${\rm M}_{jj}>1.5$~TeV,  or,  
\item $|\Delta\eta_{jj}|>6$ and ${\rm M}_{jj}>2$~TeV.
\end{enumerate} 
\end{enumerate}

\begin{figure}[!b]
  \includegraphics[width=0.5\textwidth]{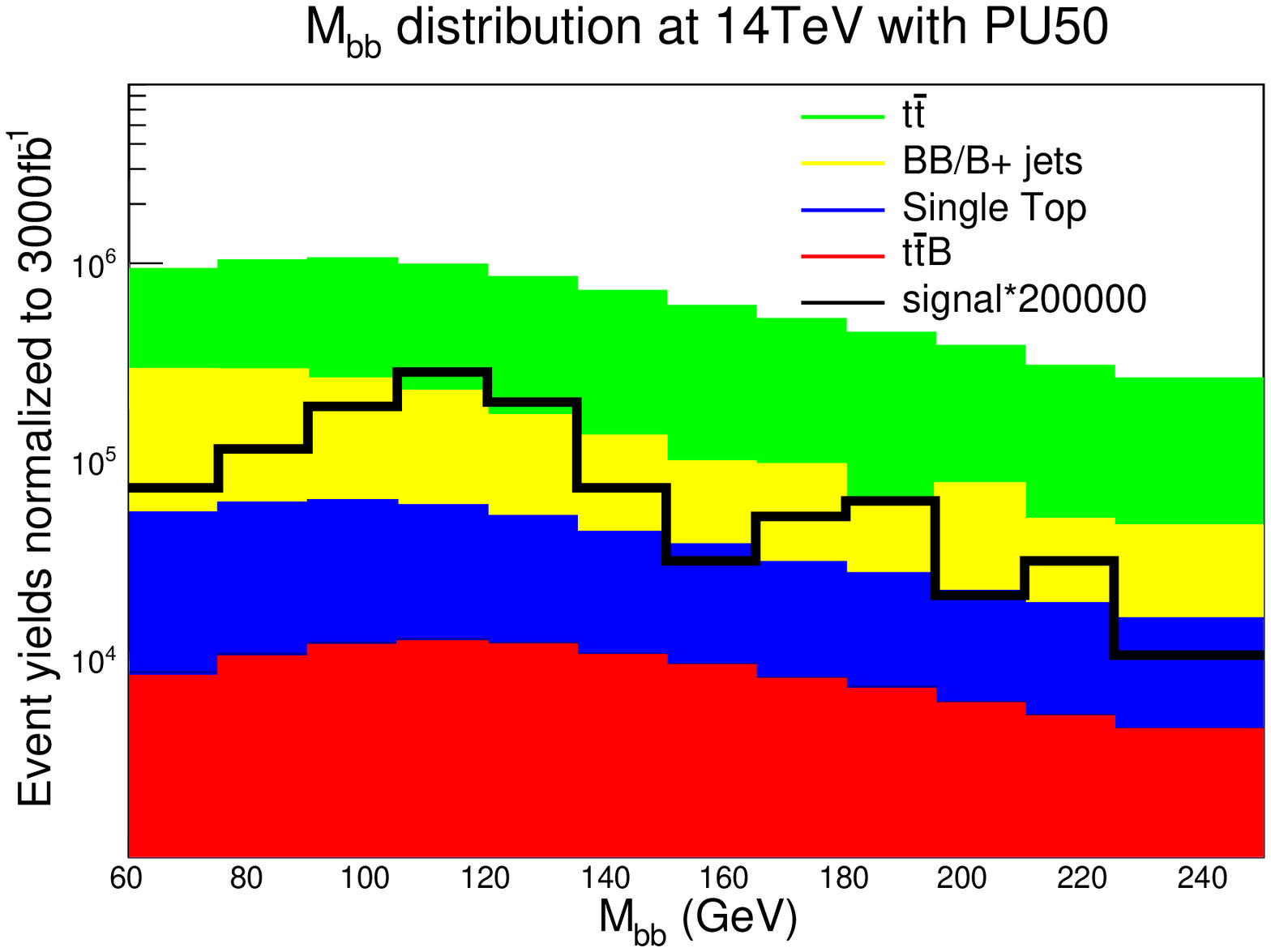}
  \hfill
  \includegraphics[width=0.5\textwidth]{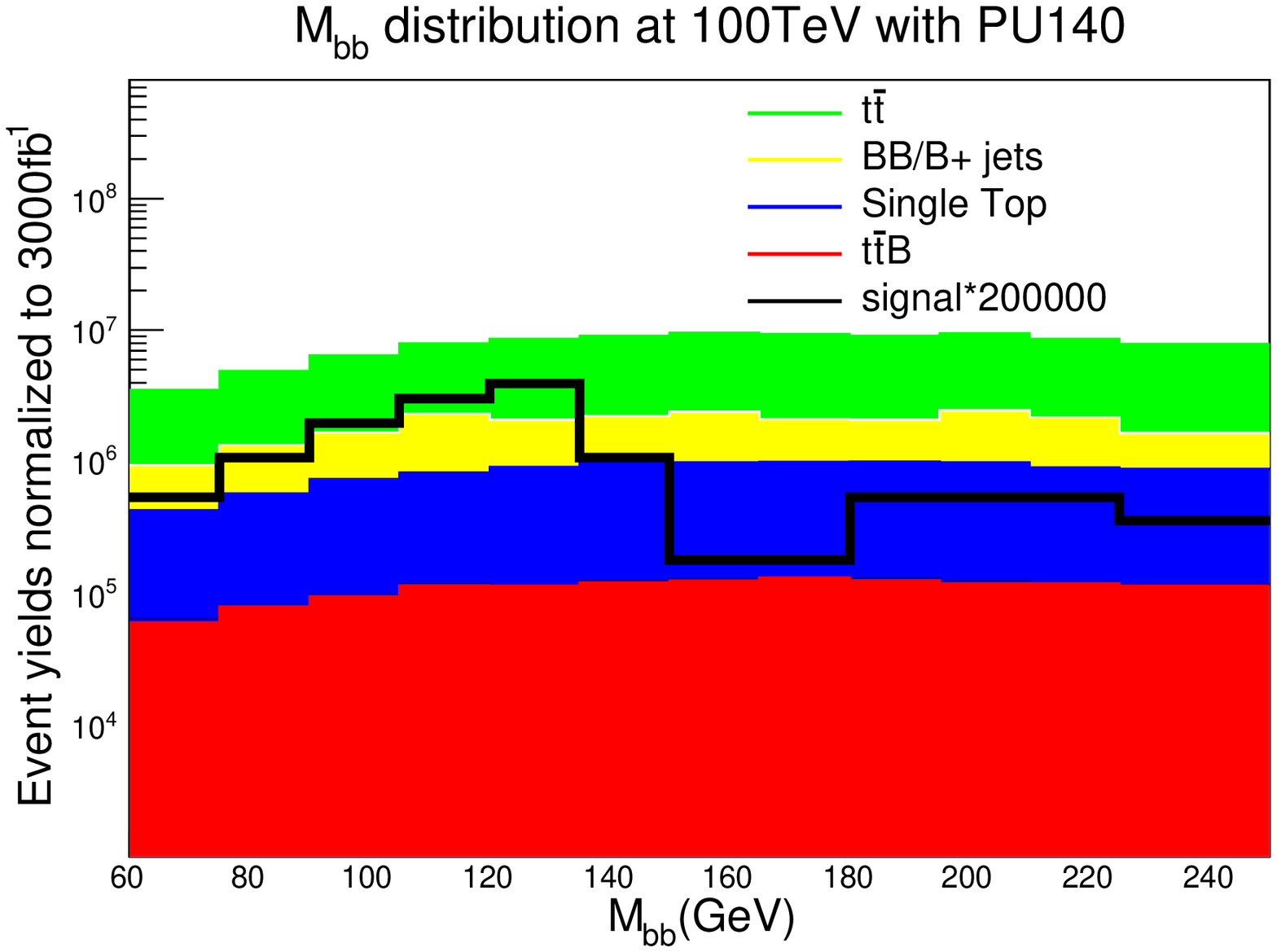}
  \caption{\label{fig:mbb} ${\rm M}_{bb}$ distributions of signal and background  for the HL-LHC at 14 TeV with 50 pileup scenario and 100 TeV with 140 pileup scenario.}
\end{figure}

Fig.~\ref{fig:mbb} shows the ${\rm M}_{bb}$ distributions of signal and background after cuts 1.)-6.) have been applied for the HL-LHC 14 TeV and 100 TeV pp collider including pileup.  One can see that although the Higgs peak can be reconstructed around 120-125 GeV for the signal, it is considerably washed out due to pileup and mistag effects. Thus we decide to choose a wide mass window in 7.) as listed above.

A cut flow for our analysis can be found in Tab.~\ref{tabcut}, which gives results for $\sqrt{s}=14$~TeV PU50. Each number represents the efficiency passing that single step's selection.  One can clearly see the power of VBF selections which can suppress backgrounds by more than two orders of magnitudes than signal.

\begin{table*}[t!]
\centering
\begin{tabular}{|c||c|c|c|c|c|c|c|c|c||}
\hline
$\rm{Cut\ Flow \ Table}$                 &1.)      &2.)         & 3.)      & 4.)       &  5.) &6.)  &7.)  &8. A) &8. B)               \\\hline
$t\bar{t}$                        &0.02\%     &  26.1\%      &99.9\%        &81.1\%        &92.8\%   &65.5\% &19.5\% &0.01\%  &0.01\%          \\\hline
$t\bar{t}+B$                      &0.49\%     &  48\%    &  99.9\%      &  91.8\%     &  90.3\%      &87.5\% &22.1\% &0.3\%    &0.02\%            \\\hline
Single Top                        &0.01\%     &  12.4\%    &  99.9\%       &  88.8\%     &  87.3\%       &81.4\% &23.5\% &0.8\%  &0.48\%         \\\hline
$B/BB$+ jets                      &0.03\%     &  0.9\%     & 100\%       &  86.4\%      &  91.3\%      &88.8\% &16.9\% &0.03\%   &0.01\%         \\\hline \hline
Signal                            &2.83\%     &  25.2\%     &   100\%     &  87.4\%      &  92\%      &89.6\% &39.8\% &34.1\%   &17.1\%   \\\hline
Signal ($2\times g_{\rm WWHH}$)   &4.11\%     &  20\%     &   100\%     &  92.7\%      &  96.7\%       &97.3\% &46.5\% &40.3\%  &25.4\%  \\\hline
Signal ($3\times g_{\rm WWHH}$)   &4.38\%     &  23.4\%    &  100\%     &  98\%    &  99\%     &98\% &40.5\% &32.9\%    &20.3\%   \\\hline \hline
Signal ($5 \times \lambda_{\rm HHH}$)  &2.91\%  &   24.6\%     &  100\%       &  93.7\%      &  91\%       &90.1\% &34.5\% &34.2\%  &23.7\%     \\\hline
\end{tabular}
\caption{\label{tabcut} Cut chain table for backgrounds and signals at the LHC with $\sqrt{s}=14$~TeV in the 50 pileup scenario.}
\end{table*}

\subsection{Numerical Results at HL-LHC}
\label{ana}

\qquad In Tab.~\ref{tab}, we show the signal and background yields at the HL-LHC with $\sqrt{s}=14$~TeV and integrated luminosity of $3000\fbinv$, after the selection cuts as listed in Sec.~\ref{Sim}. Numbers are provided for both no pileup and mean 50 pileup scenarios. The largest background contribution results from $t\bar{t}+\B$.  The remaining contributions are all found to be small. This applies to the Snowmass $\B$+jets, our produced ${\rm W}^\pm {\rm W}^\pm$ + QCD jets, and the VBS ${\rm W}^\pm {\rm W}^\pm$ (or ${\rm WZ}$) + jets contributions.

\begin{table*}[h!]
\centering
\begin{tabular}{|c||c|c||c|c||}
\hline
\multirow{2}{*}{Processes} & \multicolumn{2}{|c||}{ $\Delta\eta_{jj}>5$ and ${\rm M}_{jj}>1.5$~TeV } & \multicolumn{2}{c||}{ $\Delta\eta_{jj}>6$ and ${\rm M}_{jj}>2$~TeV }  \\ \cline{2-5} 
& {\tiny($A$) No-PU } & {\tiny($B$)  PU50}    
& {\tiny($A$) No-PU } & {\tiny($B$)  PU50} \\\hline 
$t\bar{t}$             &  0.0      &  0.86      &  0.0      &  0.86                 \\\hline 
$t\bar{t}+\B$          &   10.38     &  13.5   &  2.79      &  1.13                      \\\hline 
Single Top             &  0.156      &  7.2     &  0.06      &  4.4                  \\\hline 
$\B/\B\B$+ jets          &     0.89       & 0.07       &  0.0      &  0.03                  \\\hline 
{\bf total bkg}  & 11.43  &  21.6 &  2.85      &  6.42  \\\hline\hline 
Signal                   &  0.52          &   0.73   &  0.1      &  0.37            \\\hline 
Signal ($2\times g_{\rm WWHH}$)                  &    5.61        &   6.89   &  3.8      &  4.3           \\\hline 
Signal ($3\times g_{\rm WWHH}$)                  &  22.04          &  22.03    &  11.87      &  13.56            \\\hline \hline
Signal ($5 \times \lambda_{\rm HHH}$)                  &   1.1         &  0.8   &  0.5      &  0.8              \\\hline
\end{tabular}
\caption{\label{tab} Yields for backgrounds and signals at the LHC with $\sqrt{s}=14$~TeV and integrated luminosity of $3000\fbinv$. }
\end{table*}

One can see that our signal is not sensitive to the rescalings of trilinear Higgs coupling $\lambda_{\rm HHH}$, while there is sensitivity to $g_{\rm WWHH}$.  With an integrated luminosity of $3000\fbinv$ at the 14~TeV LHC, we expect that the gauge-Higgs quartic coupling $g_{\rm WWHH}$ can be constrained to be smaller than $\sim 2$ -- $2.5$ times of SM value at 95\% confidence level~(CL) 
\begin{equation}
\kappa_{\text{WWHH}}={g_{\text{WWHH}}\over g^{\text{SM}}_{\text{WWHH}}} = 1^{+1.2 (1.4) }_{-1}\,,
\end{equation}
without (with) pileup effects included. 
 
\begin{figure}[!t]
  \centering
  \includegraphics[width=0.6\textwidth]{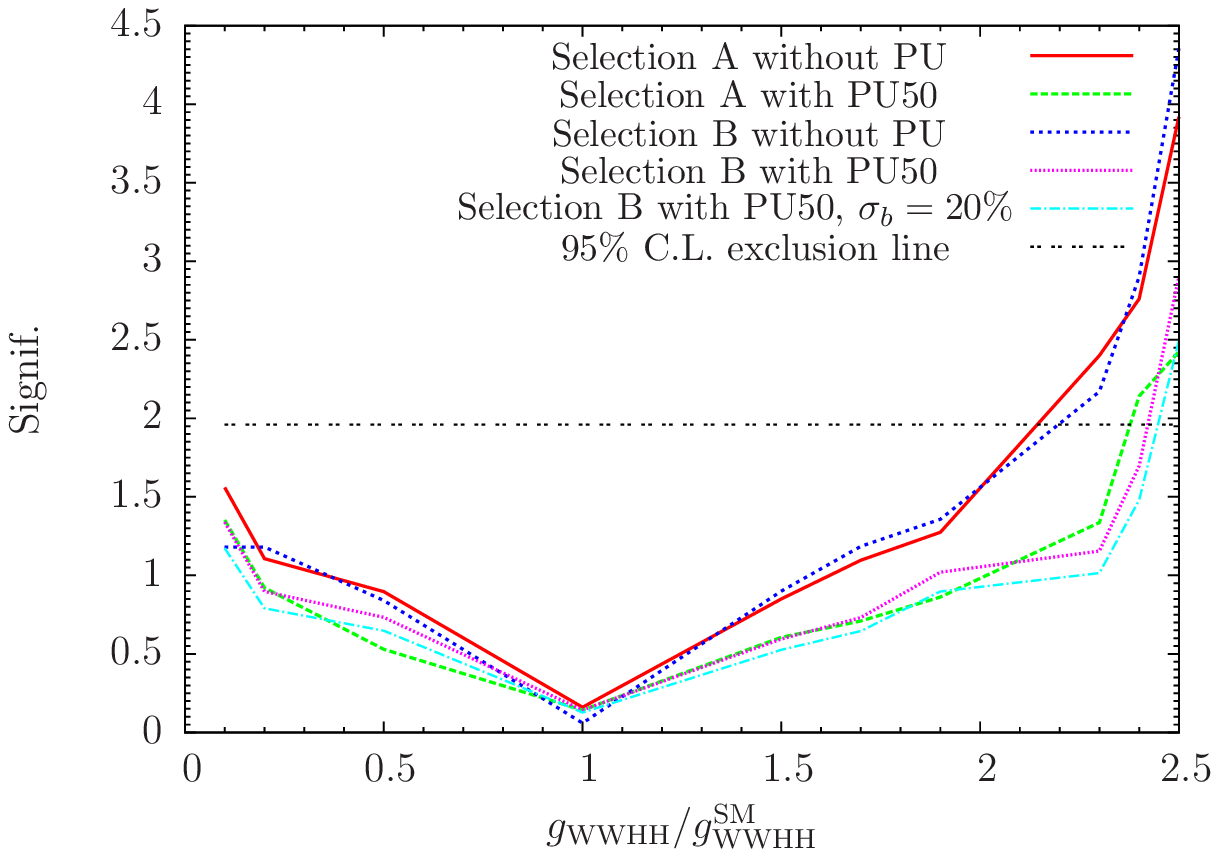}
  \caption{\label{fig:sig}  Dependence of the significance of Eq.~\eqref{stat} on $g_{\rm WWHH}$, at the 14~TeV HL-LHC with an integrated luminosity of $3000\fbinv$.}
\end{figure}

The significance distribution that underpins this result is shown in Fig.~\ref{fig:sig} (see also~\cite{atlas}), and calculated using
\begin{eqnarray}\label{stat}
\sigma = \sqrt{2 \ln(Q)}~, \quad\quad \text{} Q = (1 + N_s/N_b)^{N_{obs}} \exp(-N_s)~,
\end{eqnarray}
which corresponds to $\sqrt{2\ln [{\cal L}(S+B)/{\cal L}(B)]}$. ${\cal L}$ symbolizes the Poisson likelihood: $N_b$ is the total background yield including also the SM VBS ${\rm W}^\pm {\rm W}^\pm {\rm H}$  prediction, while $N_s$ is the signal yield defined as the excess of the signal with non-SM $g_{\rm WWHH}$ over the SM one.  $\sigma$ is related to log likelihood ratio, and a value of 1.96 corresponds to the 95\% confidence level exclusion limit in the case of only one degree of freedom.  

We have also compared scenarios with cuts 9.)~($A$) and 9.)~($B$), with and without pileup.  For the No-PU case, the more stringent cut of 9.)~(B) yields a better performance. However, pileup significantly impacts both options. As mentioned above, we include an additional sensitivity projection to Fig.~\ref{fig:sig} to be compared to 9.)~($B$), which includes the effect of 20\% systematics on background yields (we follow the procedure as suggested in~\cite{0702156}). The sensitivities do change only slightly, as the results are dominated by statistical errors.

\section{VBS ${\rm W}^\pm {\rm W}^\pm {\rm H}$ production at a at 100~TeV pp-collider}
\label{sec:100tev}
\subsection{Event Simulation and Selection}
\label{Sim100}
For the 100~TeV analysis we largely follow the cut scenario described in the above Sec.~\ref{Sim}. However, we include some modifications which optimize the cut flow for the more energetic final states compared to the HL-LHC: (1) the lepton requirement is tightened to $P_{T\,l} \geq 50$~GeV, and (2) selection cuts  8.) are changed to \begin{alignat*}{5}
\text{($A^{\ast}$)} &\quad && |\Delta\eta_{jj}|>6,{\text{~and~}}{\rm M}_{jj}>2~\text{TeV}\,, \\
\text{($B^{\ast}$)} &\quad && |\Delta\eta_{jj}|>7,{\text{~and~}}{\rm M}_{jj}>2~\text{TeV}\,.
\end{alignat*}
As for 14 TeV, we again include the impact of pileup to our discussion of results. As pileup will increase at 100 TeV compared to the 14 TeV collisions, we concentrate on the No-PU and PU140 scenarios.

\begin{table*}[t!]
\centering
\begin{tabular}{|c||c|c||c|c||}
\hline
\multirow{2}{*}{Processes} & \multicolumn{2}{|c||}{ $\Delta\eta_{jj}>6$ and ${\rm M}_{jj}>2$~TeV } & \multicolumn{2}{c||}{ $\Delta\eta_{jj}>7$ and ${\rm M}_{jj}>2~TeV$ }  \\ \cline{2-5}
& {\tiny($A^{\ast}$) No-PU } & {\tiny($B^{\ast}$)  PU140}
& {\tiny($A^{\ast}$) No-PU } & {\tiny($B^{\ast}$)  PU140} \\\hline
$t\bar{t}$             &  121      &  17240      &  0      &  975                 \\\hline
$t\bar{t}+\B$          &   618     &  1432   &  207      &  314                      \\\hline
Single Top             &  113      &  6157     &  0      &  270                  \\\hline
$\B/\B\B$+jets          &     0.96       & 239       &  0      &  0                 \\\hline
{\bf total bkg}  & 853  &  25068 &  207      &  1559  \\\hline\hline
Signal                   &  15.2          &   13.4   &  8.94      &  8.05            \\\hline
Signal ($2\times g_{\rm WWHH}$)                  &    927        &   948   &  625      &  689           \\\hline
Signal ($3\times g_{\rm WWHH}$)                  &  3457          &  3553    &  2785      &  2881            \\\hline \hline
Signal ($5 \times \lambda_{\rm HHH}$)                  &   0.75         &  0.69   &  0.27      &  0.48              \\\hline
\end{tabular}
\caption{\label{tab02} Signal and background yields at the 100~TeV pp-collider with an integrated luminosity of $3000\fbinv$.}
\end{table*}

\subsection{Numerical Results}
\label{ana100}

\begin{figure}[!t]
  \includegraphics[width=0.5\textwidth]{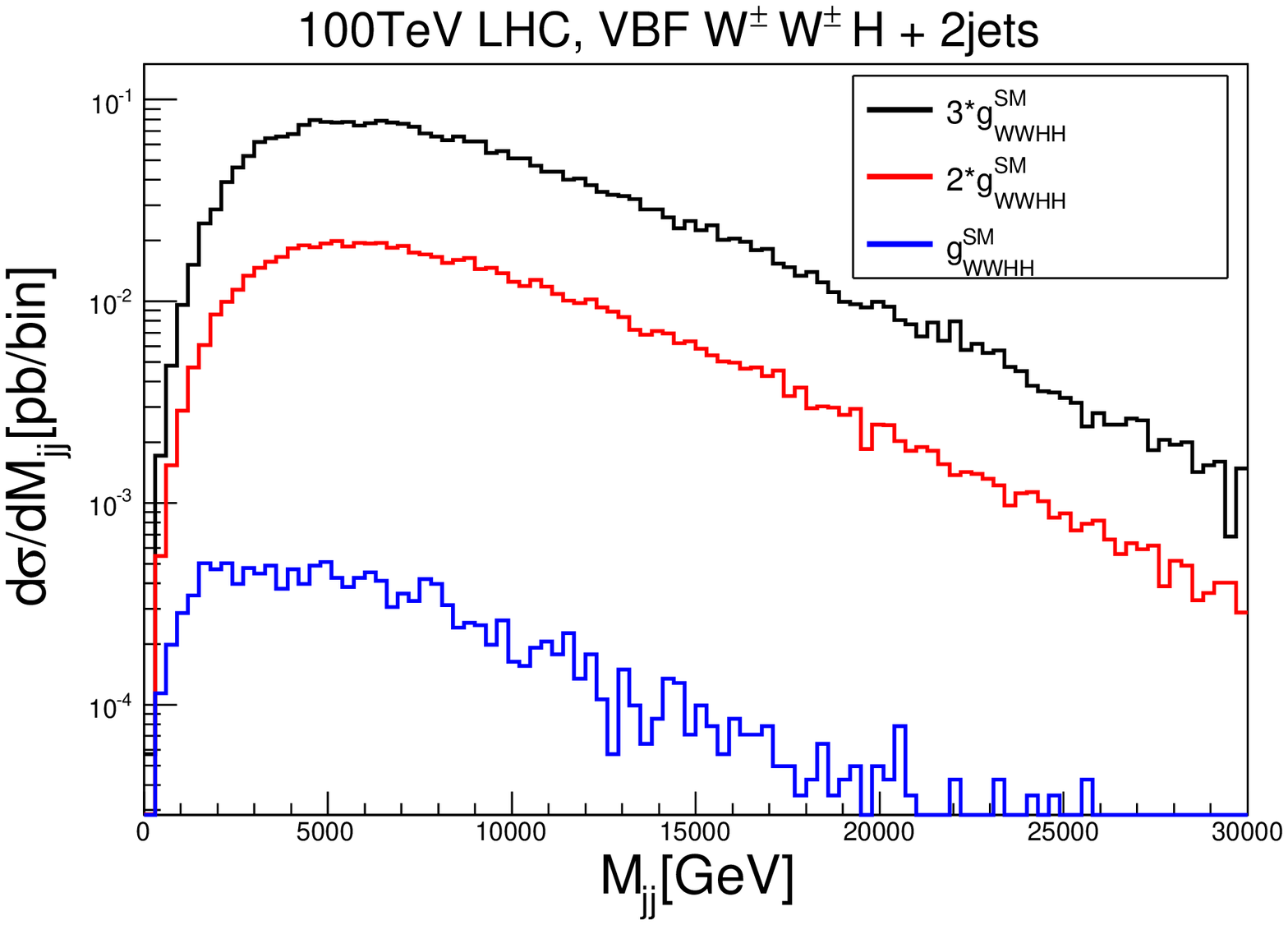}
    \hfill
  \includegraphics[width=0.5\textwidth]{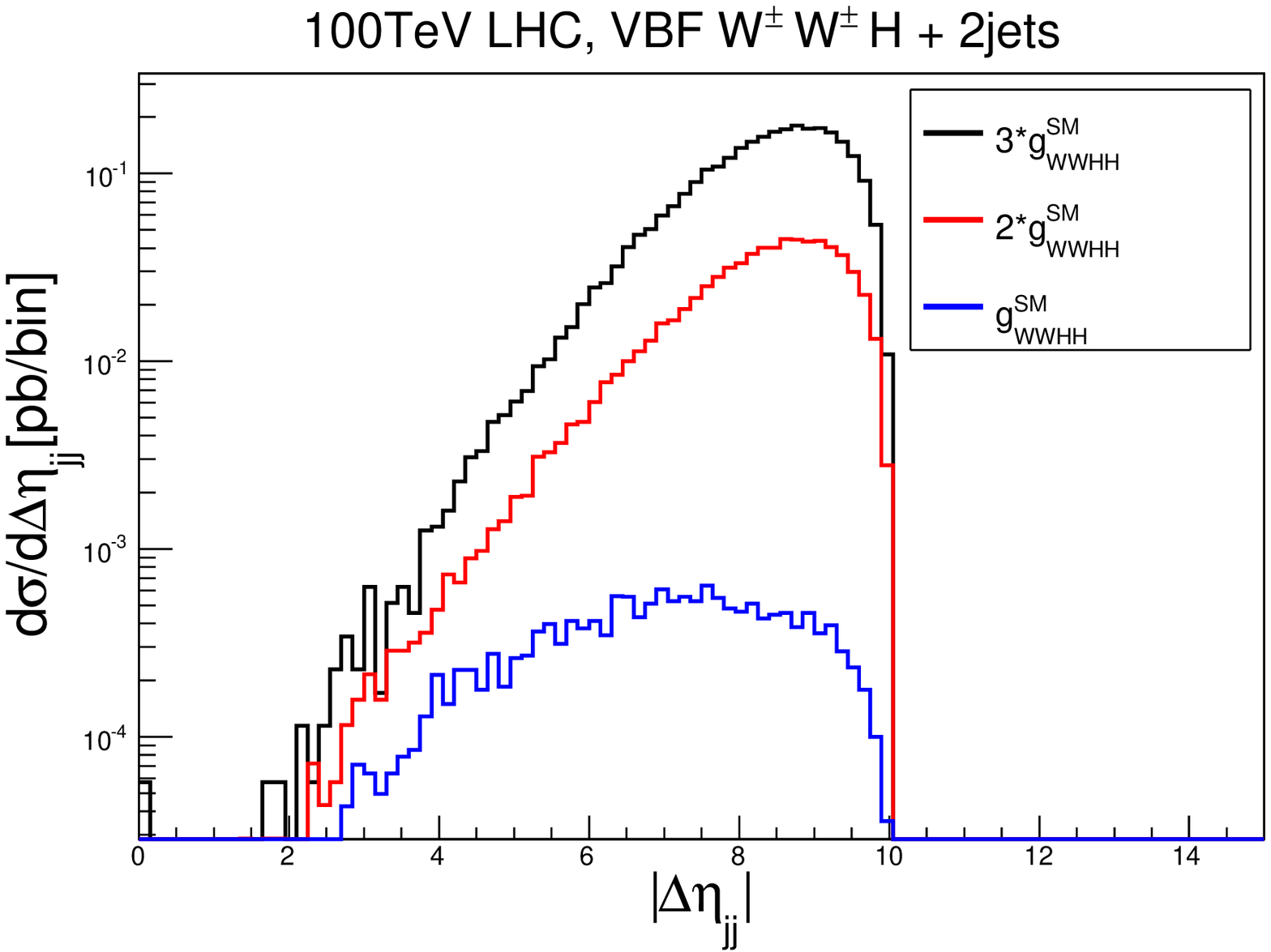}
  \caption{\label{fig:parton02} ${\rm M}_{jj}$ and $\Delta\eta_{jj}$ distributions for ${\rm W}^\pm {\rm W}^\pm H$ productions at the 100~TeV pp-collider, in the SM and for varied $g_{\rm WWHH}$ cases. }
\end{figure}

In Fig.~\ref{fig:parton02}, we show the VBS jets' invariant mass ${\rm M}_{jj}$ and their pseudo-rapidity gap $|\Delta\eta_{jj}|$ at parton level for three different $g_{\rm WWHH}$ values. One can see that both distributions are shifted to higher values compared with the HL-LHC case. Tab.~\ref{tab02} provides signal and background yields after the full selection at a 100~TeV pp-collider for an integrated luminosity of $3000\fbinv$. Note that for the 100~TeV center-of-mass energy, $t\bar{t}$ becomes the most dominating background, as non-prompt leptons from hadron decays can be energetic now and pass respective selections.

\begin{figure}[!t]
  \centering
  \includegraphics[width=0.6\textwidth]{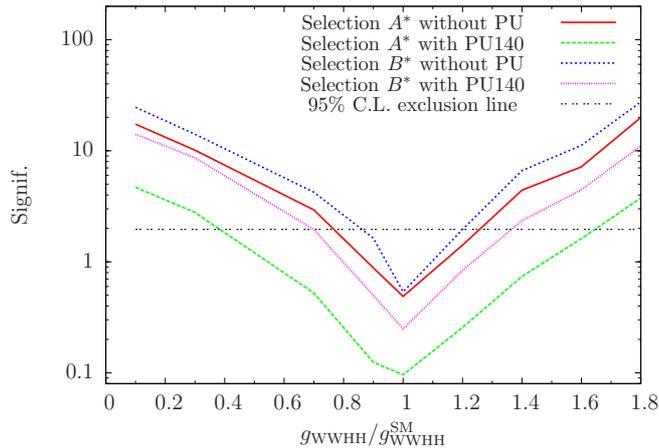}
  \caption{\label{fig:sig100TeV} Significance dependence on $g_{\rm WWHH}$, at the 100~TeV pp-collider with an integrated luminosity of $3000\fbinv$. Again the significance follows Eq.~\eqref{stat}.}
\end{figure}

The projected sensitivity to $g_{\rm WWHH}$ is shown in Fig.~\ref{fig:sig100TeV}, from which one can see that $g_{\rm WWHH}$ can
now be further constrained to be 
\begin{equation}
\kappa_{\text{WWHH}}={g_{\text{WWHH}}\over g^{\text{SM}}_{\text{WWHH}}} = 1^{+0.2 (0.4) }_{-0.1 (0.3)}\,,
\end{equation}
without (with) pileup effects included, i.e. within $\sim 20-30\%$ around SM prediction at 95\% CL. This is a significant improvement over the LHC projection. We have also compared scenarios with cuts 8.)~($A^{\ast}$)~and 8.)~($B^{\ast}$), and with or without pileup. For both these cases, the more stringent option 8.)~($B^{\ast}$) gives better performance than 8.)~($A^{\ast}$), owing to the high energetic final states that can be accessed at the 100 TeV machine.

\section{Summary and Conclusions}
\label{talk}
\qquad VBS ${\rm W}^\pm {\rm W}^\pm {\rm H}$+2 jet production is a so far unconsidered process with the potential to add sensitivity to the current Higgs characterization program. The  same-sign leptonic final state is particularly clean on top of good additional background suppression handles motivated from VBS Higgs+2~jet production. Our results show that at the high luminosity LHC with a target of $3000\fbinv$ we can expect a similar sensitivity to the quartic ${\rm WWHH}$ coupling as provided by VBS ${\rm HH}$ production, for which we expect $\kappa_{VVHH}\simeq 1.6$~\cite{Dolan:2015zja}.\footnote{It is worth mentioning that Ref.~\cite{Dolan:2015zja} did not include expected pileup or detector response effects.} Therefore, VBS ${\rm W}^\pm {\rm W}^\pm {\rm H}$+2 jet can assist in disentangling the individual contributions of the quartic gauge-Higgs vertices. In the search region selected by a maximum background rejection, modifications of the trilinear Higgs coupling have no significant impact on the signal yield. Adapting our study to the 100~TeV pp-collider we find that the $g_{\rm WWHH}$ coupling can be constrained significantly better within $\sim 20\%$ around SM prediction at 95\% CL for a comparable luminosity as the HL-LHC. Therefore, this process and its impact can be considered as another motivation to push the high energy frontier.

\acknowledgments
This work is supported in part by the National Natural Science Foundation of China, under Grants No. 11475190 and No. 11575005, and by the CAS Center for Excellence in Particle Physics (CCEPP). MS is supported in part by the European Commission through the ``HiggsTools'' Initial Training Network PITN-GA-2012-316704. 


\appendix


\end{document}